\documentclass[prl,twocolumn,showpacs,amsfonts,amsmath,floatfix]{revtex4}
\usepackage{graphicx}
\newcommand{\Journal}[4]{#1 {\bf #2}, #3 (#4)}
\newcommand{\PR}{Phys. Rev.}
\newcommand{\PRL}{Phys. Rev. Lett.}
\newcommand{\PRA}{Phys. Rev. A}
\newcommand{\JMP}{J. Math. Phys.}

\newcommand{\Science}{Science}

\newcommand{\PLA}{Phys. Lett. A}
\begin{document}
\title {Static and Dynamic Properties of Trapped Fermionic Tonks-Girardeau Gases}
\author{M. D. Girardeau}
\email{girardeau@optics.arizona.edu}
\author{E. M. Wright }
\email{ewan.wright@optics.arizona.edu}
\affiliation{Optical
Sciences Center, University of Arizona,
Tucson, AZ 85721}
\date{\today}
\begin{abstract}
We investigate some exact static and dynamic properties of
one-dimensional fermionic Tonks-Girardeau gases in tight de
Broglie waveguides with attractive p-wave interactions induced by
a Feshbach resonance. A closed form solution for the one-body
density matrix for harmonic trapping is analyzed in
terms of its natural orbitals, with the surprising result that for
odd, but not for even, numbers of fermions the maximally occupied natural 
orbital coincides with the ground harmonic oscillator orbital and has
the maximally allowed fermionic occupancy of unity. The exact 
dynamics of the trapped gas following turnoff of the p-wave
interactions are explored.

\end{abstract}
\maketitle
The spin-statistics theorem, according to which identical
particles with integer spin are bosons whereas those with
half-integer spin are fermions, breaks down if the particles are
confined to one or two dimensions. Realization of this fact had
its origin when it was shown by one of us \cite{Gir60} that the
many-body problem of hard-sphere bosons in one dimension can be
mapped exactly onto that of an ideal Fermi gas via a Fermi-Bose
(FB) mapping. It is now known that this ``Fermi-Bose duality'' is
a very general property of identical particles in one-dimension
(1D). In recent years this subject has become highly
relevant through experiments on ultracold atomic vapors in atom
waveguides
\cite{Key,Muller,Thy,Bongs,Denschlag,Greiner,GorVogLea01,Dek00,Par04,MorStoKohEss03,Sch01,Tol04,Kin04}.
Exploration of these systems is facilitated by tunability of their
interactions by external magnetic fields via Feshbach resonances
\cite{Rob01}. In fermionic atoms in the same spin state, s-wave
scattering is forbidden by the exclusion principle and p-wave
interactions are usually negligible. However, they can be greatly
enhanced by Feshbach resonances, which have recently been observed
in an ultracold atomic vapor of spin-polarized fermions
\cite{RegTicBohJin03}. Additional resonances are induced by tight
transverse confinement in an atom waveguide. Of particular
interest is the regime of low temperatures and densities where
transverse oscillator modes are frozen and the dynamics is
described by an effective 1D Hamiltonian with zero-range
interactions \cite{Ols98,PetShlWal00}, a regime already reached
experimentally
\cite{Greiner,GorVogLea01,MorStoKohEss03,Tol04,Sch01,Par04,Kin04}.
Transverse modes are still \emph{virtually} excited during
collisions, leading to renormalization of the effective 1D
coupling constant via a confinement-induced resonance. This was
first shown for bosons \cite{Ols98} and recently explained in
terms of Feshbach resonances associated with bound states in
closed, virtually excited transverse oscillation channels
\cite{BerMooOls03}, and recently Granger and Blume \cite{GraBlu03}
have shown that the same phenomenon occurs in spin-polarized
fermionic vapors.

For a system of bosons with hard core repulsive interactions, a
Tonks-Girardeau (TG) gas, the FB mapping relates the ground state
for the system to the ground state of a free Fermi gas
\cite{1dsho}. We have recently pointed out
\cite{GirOls03,GirNguOls04} that in the case of the spin-aligned
Fermi gas, the generalized FB mapping
\cite{CheShi98,GraBlu03,GirOls03} can be exploited in the opposite
direction to map the ``fermionic TG gas''
\cite{GirOls03,GirNguOls04}, a spin-aligned Fermi gas with strong
1D atom-atom interactions mediated by a 3D p-wave Feshbach
resonance, to the ground state of the trapped \emph{ideal Bose}
gas.  In this Letter we study some new and exact static and
dynamic properties of fermionic TG gases.

{\it Spin-aligned Fermi gas:} A magnetically trapped, spin-aligned
atomic vapor of spin-$\frac{1}{2}$ fermionic atoms in a tight
waveguide is magnetically frozen in the spin configuration
$\uparrow_{1}\cdots\uparrow_{N}$, so the space-spin wave function
must be \emph{spatially} antisymmetric, s-wave scattering is
forbidden, and the leading interaction effects at low energies are
determined by the 3D p-wave scattering amplitude. For a 1D gas of
free spin-aligned fermions with harmonic trapping of frequency
$\omega$ the Hamiltonian is
\begin{equation}\label{eq1}
\hat{H}_{0}=\sum_{p=1}^{N}\hat{H}(x_p) = \sum_{p=1}^{N}
\left[-\frac{\hbar{^2}}{2m}\frac{\partial^2}{\partial x_{p}^{2}}
+\frac{1}{2}m\omega^{2}x_{p}^{2}\right]  ,
\end{equation}
and the $N$-fermion ground state $\psi_{F0}(x_{1},\cdots,x_{N})$
is a Slater-determinant of the $N$ lowest HO orbitals
$\phi_{n}(x)=
e^{-Q^{2}/2}H_{n}(Q)/\pi^{1/4}x_{osc}^{1/2}\sqrt{2^{n}n!}$
\cite{1dsho}, with $H_n(Q)$ the Hermite polynomials and
$Q=x/x_{osc}$ where $x_{osc}=\sqrt{\hbar/m\omega}$ is the
longitudinal oscillator width. The Fermi-Bose mapping method
\cite{Gir60} maps this spin-aligned $N$-fermion ground state to
the $N$-boson ground state of a system of 1D impenetrable bosons
(TG gas) leading to an explicit expression for the $N$-boson
ground state \cite{1dsho}.

Here we consider the opposite limit of the ``fermionic TG gas''
\cite{GirOls03,GirNguOls04}, a spin-aligned Fermi gas with strong
1D atom-atom interactions induced by a 3D p-wave Feshbach
resonance \cite{RegTicBohJin03}. Granger and Blume derived an
effective one-dimensional K-matrix for longitudinal scattering of
two spin-aligned fermions confined in a single-mode harmonic atom
waveguide \cite{GraBlu03}. It can be shown
\cite{GraBlu03,GirOls03,GirNguOls04} that in the low-energy domain
the K-matrix can be reproduced, with a relative error as small as
${\cal O}(k_{z}^3)$, by the contact condition
\begin{equation}\label{Fermi-contact}
\psi_{F}(x_{j\ell}=0+)=-\psi_{F}(x_{j\ell}=0-)
= -a_{1D}^{F}\psi_{F}^{'}(x_{j\ell}=0\pm)
\end{equation}
where the prime denotes differentiation,
\begin{eqnarray}\label{renorm}
a_{1D}^{F}&=&\frac{6V_{p}}{a_{\perp}^2}[1+12(V_{p}/a_{\perp}^3)
|\zeta(-1/2,1)|]^{-1}
\end{eqnarray}
is the one-dimensional scattering length,
$a_{\perp}=\sqrt{\hbar/\mu\omega_{\perp}}$ is the transverse
oscillator length, $V_{p}=a_{p}^{3}=-\lim_{k\to
0}\tan\delta_{p}(k)/k^3$ is the p-wave ``scattering volume''
\cite{SunEsrGre03}, $a_{p}$ is the p-wave scattering length,
$\zeta(-1/2,1)=-\zeta(3/2)/4\pi=-0.2079\ldots$ is the Hurwitz zeta
function evaluated at $(-1/2,1)$ \cite{WhiWat52}, and $\mu=m/2$ is
the reduced mass. The expression (\ref{renorm}) has a resonance at
a \emph{negative} critical value
$V_{p}^{crit}/a_{\perp}^{3}=-0.4009\cdots$. In accordance with
(\ref{Fermi-contact}), the low-energy fermionic wavefunctions are
discontinuous at contact, but left and right limits of their
derivatives coincide, and following Ref.~\cite{CheShi98} we assume
the same here. For an interatomic interaction of short but nonzero
range $x_0$, the fermionic relative wave function vanishes at
$x_{j\ell}=0$; Eq. (\ref{Fermi-contact}) refers to the $x_{0}\to
0+$ limit of the exterior wave function just outside the range of
the interaction \cite{CheShi98}, \cite{GirOls03}.

The contact condition (\ref{Fermi-contact}) is generated by the 1D
interaction pseudopotential operator \cite{GirOls04,GirNguOls04}
$\hat{v}_\text{int}^{F}=g_{1D}^{F}\sum_{1\le j<\ell\le
N}\delta^{'}(x_{j\ell}) \hat{\partial}_{j\ell}$ where
$\hat{\partial}_{j\ell}\psi=(1/2)[\partial_{x_{j}}\psi|_{x_{j}=x_{\ell
+}} -\partial_{x_{\ell}}\psi|_{x_{j}=x_{\ell -}}]$ with effective
1D coupling constant $g_{1D}^{F}=\hbar^{2}a_{1D}^{F}/\mu$
\cite{GraBlu03,MooBerOls04,GirNguOls04}. This can be compared with
the bosonic 1D coupling constant $g_{1D}^{B}=-\hbar^{2}/\mu
a_{1D}^B$ \cite{GraBlu03,GirNguOls04}. The dimensionless fermionic
coupling constant is $\gamma_{F}=-mg_{1D}^{F}n/\hbar^2$. Note that
the density $n$ is in the numerator, whereas it is in the
denominator of the bosonic analog
$\gamma_{B}=mg_{1D}^{B}/n\hbar^2$.

{\it Fermionic TG gas:} The fermionic TG gas regime is
$\gamma_{F}\gg 1$ valid in the neighborhood of the resonance where
the denominator of Eq. (\ref{renorm}) vanishes. The 1D scattering
length is invariant under the FB mapping
\cite{Gir60,GraBlu03,CheShi98,GirOls03}
$\psi_{B}=A(x_{1},\cdots,x_{N})\psi_F$ with
$A(x_{1},\cdots,x_{N})=\prod_{1\le j<\ell\le
N}\text{sgn}(x_{\ell}-x_{j})$ the ``unit antisymmetric function''
employed in the original discovery of fermionization \cite{Gir60},
so we shall henceforth indicate it by
$a_{1D}=a_{1D}^{F}=a_{1D}^{B}$. Since $V_p$ must be negative to
achieve a resonance, the fermionic TG regime with attractive
p-wave interactions $a_{1D}\to -\infty$ and $g_{1D}^{F}\to
-\infty$ is reached as $V_p$ approaches the resonance through
negative values, implying an interaction-free exterior wave
function. More generally, the spin-aligned Fermi gas maps to the
Lieb-Liniger Bose gas \cite{LieLin63} with fermionic and bosonic
coupling constants inversely related \cite{CheShi98} according to
$g_{1D}^{B}=-\hbar^{4}/\mu^{2}g_{1D}^{F}>0$. The corresponding
dimensionless coupling constants
$\gamma_{B}=mg_{1D}^{B}/n\hbar^2>0$ and
$\gamma_{F}=-mg_{1D}^{F}n/\hbar^2>0$ satisfy
$\gamma_{B}\gamma_{F}=4$ \cite{GirOls03}.

{\it Natural orbitals:} The $N$-atom ground state of the
longitudinally trapped fermionic TG gas maps to the trapped ideal
Bose gas ground state ($\gamma_{B}=0$) for which all $N$ atoms are
Bose-Einstein condensed into the lowest HO orbital $\phi_0(x)$:
$\psi_{F}(x_{1},\cdots,x_{N};t=0)=A(x_{1},\cdots,x_{N})
\prod_{j=1}^{N}\phi_{0}(x_{j})$. Since $A^{2}=1$, all properties 
expressible in terms
of $|\psi_{F}|^2$, including the density profile, are the same as
those of the trapped Bose gas ground state
$\prod_{j=1}^{N}\phi_{0}(x_{j})$. However, properties not so
expressible, such as the momentum distribution, differ
dramatically, as in the case of the usual bosonic TG gas. The
appropriate tool for studying such differences is the reduced
one-body density matrix (OBDM)
$\rho_{1}(x,x';t)=N\int\psi_{F}(x,x_{2},\cdots,x_{N};t)
\psi_{F}^{*}(x',x_{2},\cdots,x_{N};t)dx_{2}\cdots dx_N$. The normalized 
natural orbitals satisfy $\lambda_j
u_j(x)=\int_{-\infty}^{\infty} dx' \rho_1(x,x';t=0)u_j(x')$, with
eigenfunctions $u_j(x)$, $j=0,1,2,\ldots$, and eigenvalues or
orbital occupations $0\le \lambda_j \le 1$. For a
spatially uniform bosonic TG gas of length L the natural orbitals
are plane waves $u_j(x)=e^{i2\pi xj/L}/\sqrt{L}$ and the
distribution of occupations $\lambda_j$ reflects a
momentum distribution function with low-momentum peaking
and broad wings, whereas for a free Fermi gas the ground state is
a filled Fermi sea with $\lambda_j=1, j=0,1,\ldots,N-1$, all other
occupations being zero \cite{Par04,Kin04,1dsho}. Although natural
orbitals are much-used in quantum chemistry, the first application
to degenerate gases is due to Dubois and Glyde \cite{DubGly03}.

In the case of the longitudinally trapped fermionic TG gas, the
OBDM can be evaluated in closed form for all N \cite{BenErkGra04} yielding
$\rho_1(x,x';t=0)=N\phi_{0}(x)\phi_{0}(x')[F(x,x')]^{N-1}$ where
$F(x,x')=\int_{-\infty}^{\infty}\text{sgn}(x-x'')\text{sgn}(x'-x'')
\phi_{0}^{2}(x'')dx'' =1-|\text{erf}(Q')-\text{erf}(Q)|$, with
$\text{erf}(Q)$ the error function \cite{AbrSte64}. We have
numerically diagonalized the OBDM to obtain the natural orbitals
and occupations for a variety of $N$. In Figs.~\ref{fig1}(a) and
(b) we show the occupations $\lambda_j$ versus orbital number $j$
for $N=2$ and $N=3$, respectively, as representative of our
findings for even and odd numbers of atoms. (The solid line
between data points is included as an aid to the eye). For $N=2$
we see that the occupancy distribution shows a staircase structure
where $\lambda_{2q}\simeq \lambda_{2q+1}, q=0,1,2\dots$, the
occupancies being monotonically decreasing
$\lambda_q>\lambda_{q+1}$. Furthermore, the occupancy distribution
extends well beyond the Fermi sea for the corresponding free Fermi
gas shown as the dashed line, reflecting the strongly interacting
nature of the fermionic TG gas . For $N=3$ shown in
Fig.~\ref{fig1}(b) a similar staircase structure is seen with the
exception that $\lambda_0=1$, that is the natural orbital $u_0(x)$
has the maximal occupancy allowed by the exclusion principle,
and $\lambda_{2q+1}\simeq \lambda_{2q+2},
q=0,1,2\dots$.

The natural orbitals are
structurally similar to the HO orbitals and share the property
that the orbital number $j$ is equal to the number of field nodes.
To examine the properties of the maximally occupied natural
orbital we used a Rayleigh-Ritz variational procedure with a
Gaussian trial solution $u_0(x)=\exp(-x^2/2w^2)/\pi^{1/4}\sqrt{w}$
to obtain the corresponding occupancy $\lambda_0(w)$ numerically.
We have found that for odd $N$ the maximal 
occupancy obtained by varying $w$ is $\lambda_0(w=x_{osc})=1$ for
$w=x_{osc}$, so that the maximally occupied natural orbital
$u_0(x)$ coincides with the ground state $\phi_0(x)$ of the
single-particle HO. We have verified this numerical result by
proving analytically, by tedious rearrangement and summation of series, 
that $\lambda_0=1$ assuming
$u_0(x)=\phi_0(x)$ for odd $N$, an exact and surprising
result. In contrast, for even $N$ the maximum
occupancy is obtained numerically for $w=x_{osc}/2$, the occupancy
increasing with $N$ but remaining less that unity, and comparison
of the numerical results and Gaussian solution shows that it is an
approximation to the maximally occupied natural orbital, albeit a
very good one.

\begin{figure}
\includegraphics*[width=0.5\columnwidth,
height=0.5\columnwidth]{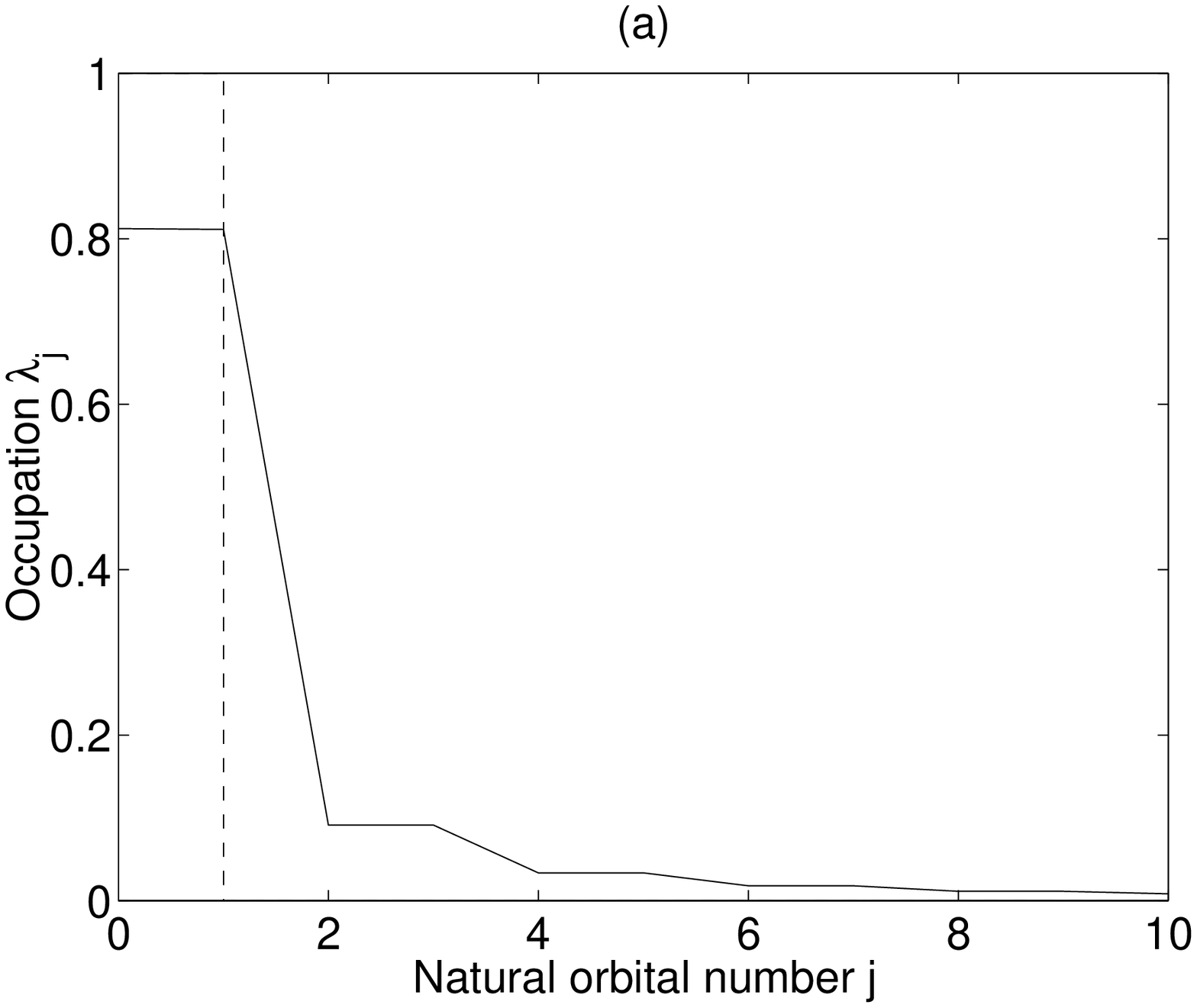}
\end{figure}

\begin{figure}
\includegraphics*[width=0.5\columnwidth,
height=0.5\columnwidth]{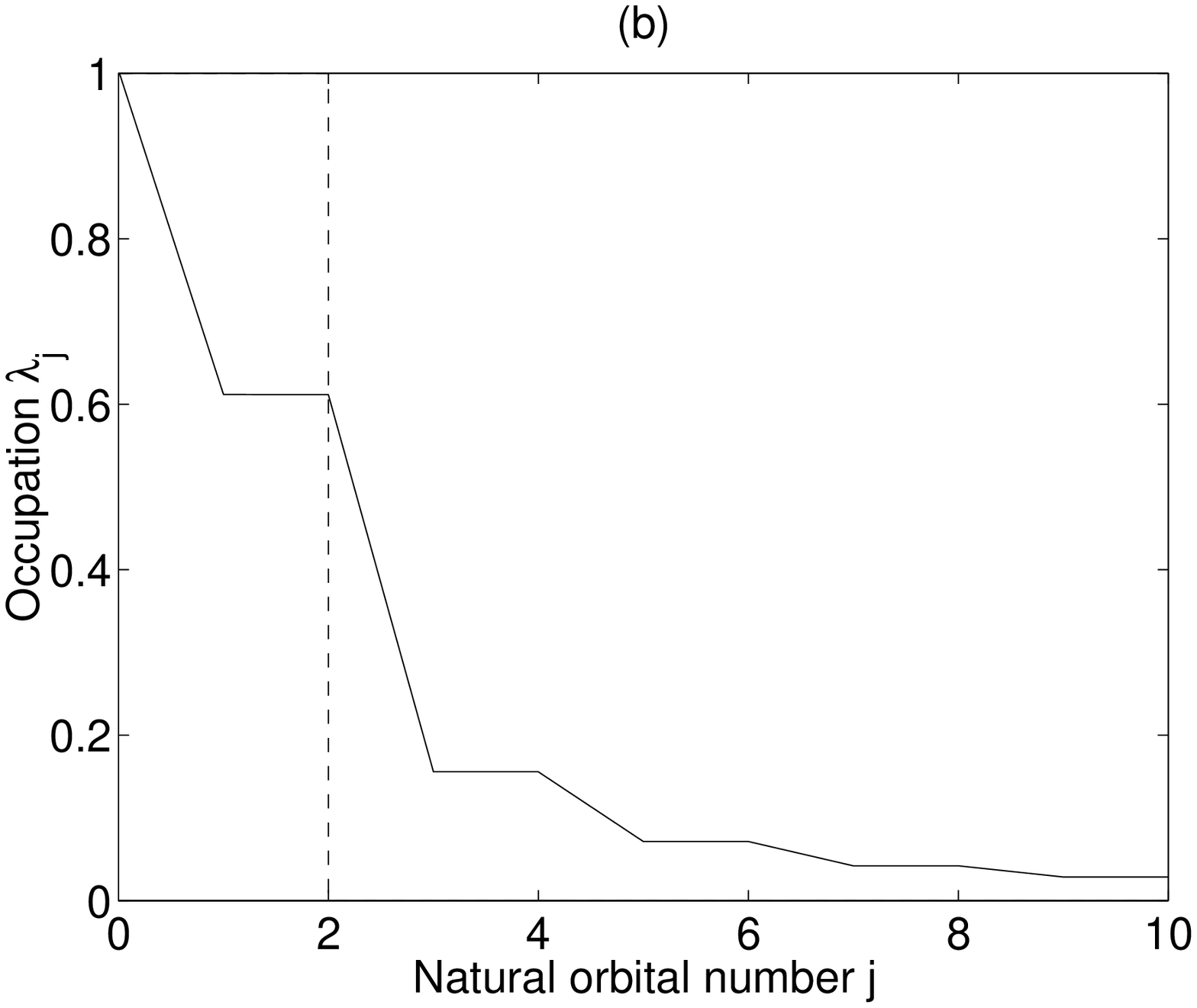} \caption{Occupation
distribution $\lambda_j$ versus natural orbital number $j$ for a)
$N=2$, and b) $N=3$. In each case the dashed line shows the
corresponding result for a free Fermi gas.} \label{fig1}
\end{figure}

{\it Dynamics following turnoff of interactions:} We next examine
the {\it exact} dynamics of the fermionic TG gas that ensue when
the interactions are suddenly turned off at $t=0$ by detuning the
external magnetic field. For $t<0$ the OBDM is given by that for
the fermionic TG gas
$\rho_{1}(x,x';t<0)=\sum_j\lambda_ju_j(x)u_j(x')$. For later
times, when the interactions are turned off, the evolution of the
OBDM is given by $i\hbar(\partial/\partial t)\rho_{1}(x,x';t)
=[\hat{H}(x)-\hat{H}(x')]\rho_{1}(x,x';t)$ \cite{Neg82}, where
$\hat{H}(x)$ is the single-particle HO Hamiltonian. Even though
this equation has no interactions the subsequent dynamics are a
direct consequence of the strong many-body correlations present in
the initial OBDM. Formally, the solution for the OBDM can be
written in terms of the HO orbitals that serve as natural orbitals
for the non-interacting gas
\begin{equation} \label{rho1}
\rho_{1}(x,x';t>0)=\sum_{m,n}C_{mn}\phi_n(x)\phi_m(x')e^{-i(n-m)\omega
t} ,
\end{equation}
where $C_{nm}=\sum_j\lambda_j\int dx u_n(x)\phi_j(x)\int dx
u_m(x)\phi_j(x)$, and the symmetry of the orbitals dictates that
$C_{nm}$ is non-zero only for $n$ and $m$ both even or both odd.
Thus, the OBDM evolution following the turnoff of the interactions
is periodic in time with period $T=\pi/\omega$. We have solved the
equation of motion for the OBDM numerically to obtain the exact
dynamics of the initial fermionic TG gas, and here we present
results for the evolution of the density profile
$\rho(x,t)=\rho_{1}(x,x;t)$.

\begin{figure}
\includegraphics*[width=0.5\columnwidth,
height=0.5\columnwidth]{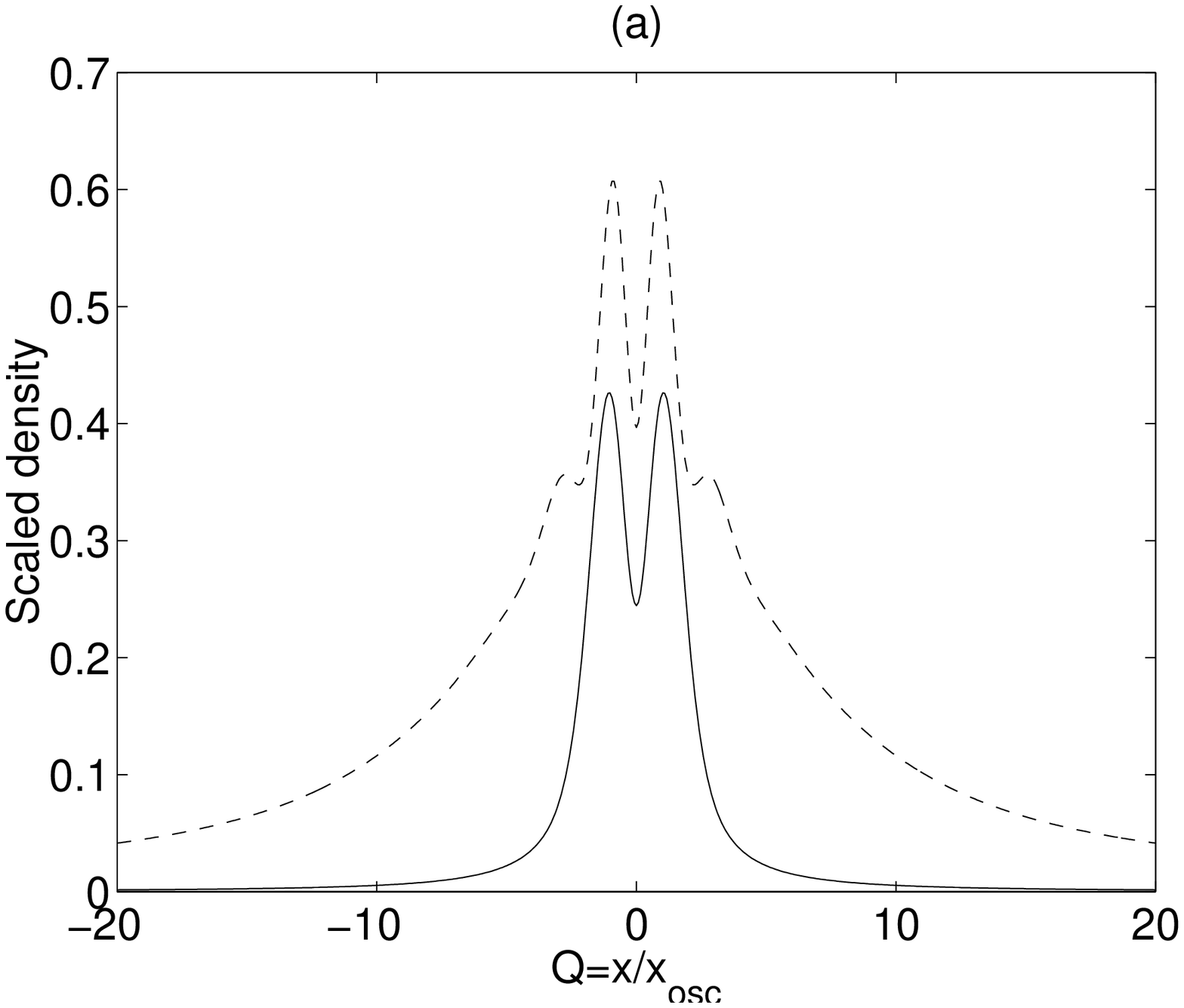}
\end{figure}

\begin{figure}
\includegraphics*[width=0.5\columnwidth,
height=0.5\columnwidth]{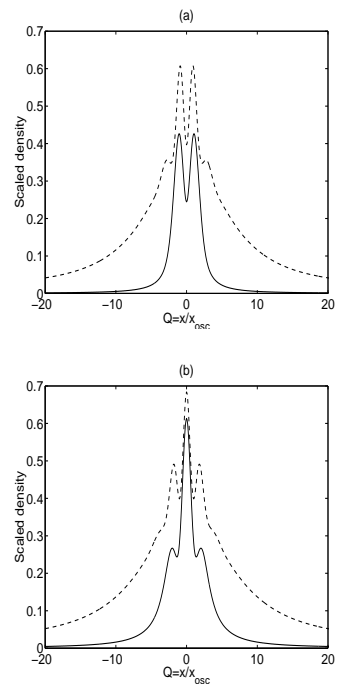} \caption{Scaled
density profile $x_{osc}\rho(Q,t=T/2)$ after half a period for a)
$N=2$ and $N=8$ (dashed line), and b) $N=3$ and $N=9$ (dashed
line).} \label{fig2}
\end{figure}

Figure~\ref{fig2}(a) shows the scaled density profile
$x_{osc}\rho(Q,t=T/2)$ after half a period, when the on-axis
density reaches its lowest value, for $N=2$ and $N=8$ (dashed
line). In each case we see that the density expands and the center
$(Q=0)$ develops a minimum. This may be intuited as follows:
Before the interactions are turned off the width of the density
profile is $\sim x_{osc}$, narrower than the width for a trapped
free Fermi gas $\sim x_{osc}\sqrt{N}$. This difference in widths
arises from the fact that the Fermi degeneracy pressure that sets
the width of the free gas in conjunction with the HO trapping
potential is countered by the attractive p-wave interactions in the
fermionic TG gas. However, when the interactions are turned off
the Fermi degeneracy pressure, which increases with $N$, acts to
broaden the density profile. For times $t>T/2$ the density profile
begins to contract again under the action of the harmonic single
particle potential, the initial condition being recovered at
$t=T$.

Figure~\ref{fig2}(b) shows the corresponding results for $N=3$ and
$N=9$ (dashed line), and similar broadening is seen at $t=T/2$. In
this case, however, a density maximum is maintained at the center
of the trap. The persistence of an on-axis density peak for odd
$N$ may be understood as follows: For odd $N$ the particles are
symmetrically placed around the center of the trap with one
particle at the center. When the interactions are turned off the
particles displaced from the center will move outwards in reaction
to the Fermi degeneracy pressure.  However, the center particle,
which is absent for even $N$, will be pinned at the center by the
equal but opposite Fermi degeneracy pressures acting on it due to
the displaced particles, and this maintains the on-axis density
peak for odd $N$. The persistence of a central peak
may also be related to the fact that for odd $N$ the maximally
occupied natural orbital $u_0(x)$ for the fermionic TG gas with
interactions is identical to the HO orbital $\phi_0(x)$ which
serves as a natural orbital with no interactions.  Thus, the
scaled density profile associated with this maximally occupied
orbital is given by
$x_{osc}\rho_0(Q)=C_{00}x_{osc}\phi_0^2(Q)=\exp(-Q^2)/\sqrt{\pi}$,
which has a maximum on-axis, integrates to unity and represents
the centrally pinned particle, and persists after the interactions
are turned off. For the $N=3$ simulation in Fig.~\ref{fig2}(b)
$x_{osc}\rho(0,T/2)\simeq 0.6$, whereas $x_{osc}\rho_0(0)=0.56$,
showing that the central peak is largely accounted for by the
maximally occupied natural orbital. For larger odd $N$ the central
maximum has a larger contribution from higher natural
orbitals but the peak
still persists; see the dashed line in Fig.~\ref{fig2}(b).

In summary, the OBDM of a harmonically trapped fermionic TG gas has been 
analyzed in terms of its natural orbitals
and occupations. For an odd number of atoms the maximally
occupied natural orbital was found to coincide with the
single-particle HO ground state. Due to the rarity of exact
results for strongly interacting systems it is important to
suggest observable consequences. We have explored the exact dynamics
following turnoff of the p-wave interactions, finding that
for odd $N$ the on-axis density remains a maximum as the cloud expands due to 
Fermi degeneracy pressure, in contrast to even $N$ where the density
develops a minimum. 

\begin{acknowledgments}
This work was initiated at the ITAMP, Harvard-Smithsonian Center
for Astrophysics, in part supported by the National Science
Foundation, and at the Benasque, Spain 2002 workshop Physics of
Ultracold Dilute Atomic Gases. It also benefited from the Quantum
Gases program at the Kavli Institute for Theoretical Physics,
University of California, Santa Barbara, supported in part by NSF
grant PH99-0794. The work of MDG is supported by the Office of
Naval Research grant N00014-99-1-0806 through a subcontract from
the University of Southern California.
\end{acknowledgments}
\end{document}